\documentclass{elsart}
\usepackage{times}
\usepackage{epsfig}
\usepackage{lineno}
\usepackage{marvosym}
\usepackage{amsmath}
\usepackage{amsfonts}
\usepackage{amssymb}
\begin{document}
\bibliographystyle{roman}

\journal{Nuclear Instruments and Methods A}

\def\hb{\hfill\break}
\def\MeV{\rm MeV}
\def\GeV{\rm GeV}
\def\TeV{\rm TeV}

\def\m{\rm m}
\def\cm{\rm cm}
\def\mm{\rm mm}
\def\lam{$\lambda_{\rm int}$}
\def\rad{$X_0$}
 
\def\NIM{Nucl. Instr. and Meth.~}
\def\ieee {{IEEE Trans. Nucl. Sci.~}}
\def\prl{Phys. Rev. Lett.~}
\def\RMP{Rev. Mod. Phys.~}

\def\etal{{\it et al.}}
\def\eg{{\it e.g.,~}}
\def\ie{{\it i.e.,~}}
\def\cf{{\it cf.~}}
\def\etc{{\it etc.~}}
\def\vs{{\it vs.~}}

\hyphenation{ca-lo-ri-me-ter}
\hyphenation{ca-lo-ri-me-ters}
\hyphenation{Brems-strah-lung}

\begin{frontmatter}
\title{Tests of a dual-readout fiber calorimeter with SiPM light sensors}

\author{M.~Antonello$^a$, M.~Caccia$^a$, M. Cascella$^b$, M.~Dunser$^c$,}
\author{R. Ferrari$^d$, S.~Franchino$^e$, G.~Gaudio$^d$, K.~Hall$^f$, J. Hauptman$^f$,} 
\author{H.~Jo$^g$, K.~Kang$^g$, B.~Kim$^g$, S. Lee$^g$, G.~Lerner$^h$, L.~Pezzotti$^i$,}
\author{R.~Santoro$^a$, I.~Vivarelli$^h$, R.~Ye$^g$ and R. Wigmans$^{j,}$\thanksref{Corres}} 

\address{$^a$ Universit\`a degli Studi dell' Insubria, Como, and INFN Sezione di Milano, Italy\\
$^b$ University College, London, UK\\
$^c$ CERN, Gen\`eve, Switzerland\\
$^d$ INFN Sezione di Pavia, Italy\\
$^e$ Kirchhoff-Institut f\"ur Physik, Ruprecht-Karls-Universit\"at Heidelberg, Germany\\
$^f$ Iowa State University, Ames (IA), USA\\
$^g$ Kyungpook National University, Daegu, Korea\\
$^h$ Sussex University, Brighton, United Kingdom\\
$^i$ Universit\`a di Pavia, Pavia, and INFN Sezione di Pavia, Italy\\
$^j$ Texas Tech University, Lubbock (TX), USA}

\thanks[Corres]{Corresponding author.
              Email wigmans@ttu.edu, fax (+1) 806 742-1182.}
              
\vskip -7mm
\begin{abstract}
In this paper, we describe the first tests of a dual-readout fiber calorimeter in which silicon photomultipliers are used to sense the (scintillation and \v{C}erenkov) light signals.
The main challenge in this detector is implementing a design that minimizes the optical crosstalk between the two types of fibers, which are 
located very close to each other and carry light signals that differ in intensity by about a factor of 60. The experimental data, which were obtained with beams of high-energy electrons and muons as well as in lab tests, illustrate to what extent this challenge was met. The \v{C}erenkov light yield, a limiting factor for the energy resolution of this type of calorimeter, was measured to be about twice that
of the previously tested configurations based on photomultiplier tubes. The lateral profiles of electromagnetic showers were measured on a scale of millimeters from the shower axis
and significant differences were found between the profiles measured with the scintillating and the \v{C}erenkov fibers.
\vskip -5mm
\end{abstract}
\end{frontmatter}
\newpage
\section{Introduction}
\vskip -5mm
In the past 15 years, the properties of dual-readout fiber calorimetry have been extensively studied by the ACCESS, DREAM and RD52 Collaborations. A recent review of the results obtained in these studies can be found in References \cite{Wig00,rpm}. The properties of this type of calorimeter are deemed very suitable for experiments at proposed future high-energy $e^+e^-$ colliders, such as FCC \cite{fcc}, CEPC \cite{cepc} or ILC \cite{ilc}. However, the detectors tested during the generic R\&D phase need to be adapted to the practical circumstances of such experiments 
in order to make this possible. This is specifically true for the readout. In the DREAM and RD52 calorimeters, Photo Multiplier Tubes (PMT) were used to detect the signals from the two types of fibers, which needed to extend about 30 cm from the rear of the calorimeter to allow separation and bunching.
In order to make this detector more suitable for the envisaged applications, it was decided to replace this readout with a system based on Silicon PhotoMultipliers (SiPM) \cite{Ren09,Buz01,Sav00,Pie06,Gar11}.  

The use of SiPMs for reading out a sampling calorimeter was pioneered by the CALICE Collaboration \cite{calice}, which built several calorimeter modules based on scintillator strips or tiles. These scintillators were connected to wavelength shifting fibers which transported the light signals to SiPMs. A similar approach is used in the shashlyk calorimeter for the COMPASS II experiment \cite{Anf13}. The Pb/scintillating-fiber barrel calorimeter of the GlueX experiment at Jefferson Lab is read out with multi-pixel photon counters that are directly coupled to the fibers \cite{glue15,sot13}.

However, SiPMs have never before been used to detect the light signals from individual scintillating or \v{C}erenkov fibers that are the active media of a dual-readout sampling calorimeter.
These solid state, single photon sensitive sensors offer potentially important specific 
advantages for the application of such calorimeters in modern experiments at colliding-beam machines:
\begin{enumerate}
\item They offer the possibility to eliminate the forests of optical fibers that stick out at the rear end. These fiber bunches occupy precious space and act as antennas for particles that come from sources unrelated to the showers developing in the calorimeter. They may also cause oversampling of late developing showers.
\item The compact readout makes it possible to separate the calorimeter into longitudinal segments, if so desired.
\item Unlike the PMTs used until now, SiPMs can operate in a magnetic field.
\end{enumerate}
As a specific additional advantage for this particular type of calorimeter, we also mention the larger quantum efficiency for photon detection, which is important since fluctuations in the number of \v{C}erenkov photoelectrons have turned out to be a limiting factor, both for the electromagnetic (em) and hadronic energy resolutions.
There are of course also potential disadvantages, most notably the fact that SiPMs are {\sl digital} detectors and therefore prone to response non-linearity and signal saturation effects. A major challenge for this particular detector concerns the fact that the SiPMs have to read the signals from a grid of closely spaced fibers of two different kinds, where the light intensity in one type of fibers (detecting the \v{C}erenkov
light) is more than an order of magnitude smaller than that in the other fibers (detecting the scintillation light). Optical crosstalk is thus a major concern.

In this paper, we describe the results of the first beam tests of a dual-readout fiber calorimeter with SiPM readout.
These tests were focused on the mentioned crosstalk and saturation effects. We also measured the \v{C}erenkov light yield.
As a byproduct, the lateral profiles of em showers very close to the shower axis were measured. It turned out that there are significant differences between the profiles measured with the two types of signals. In Section 2, the detector and the readout are described. Section 3
deals with the experimental setup in which it was tested and the methods used to analyze the data. Experimental results are the topic of Section 4 and conclusions are given in Section 5.

\section{The detector}

\begin{figure}[htbp]
\epsfysize=5cm
\centerline{\epsffile{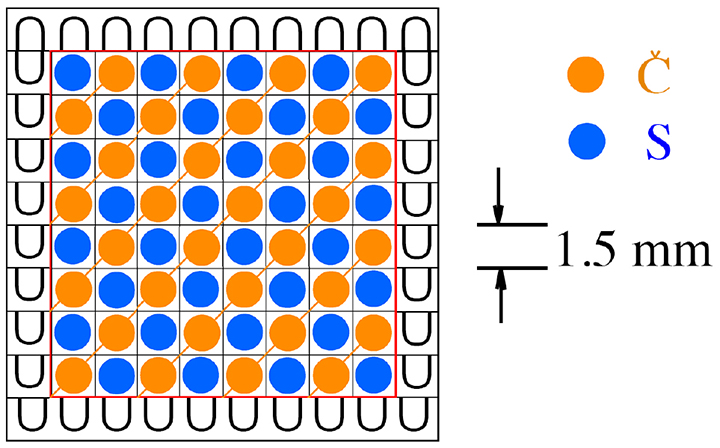}}
\caption{\footnotesize Arrangement of the 64 optical fibers in the calorimeter module. The absorber structure, made out of brass, was 112 cm long.
The scintillating (S) and \v{C}erenkov (C) fibers are indicated with different colors. For color see the online version.}
\label{module}
\end{figure}

The calorimeter used for these studies consisted of brass (Cu260). It was 112 cm long and had a lateral cross section of 15 $\times$15 mm$^2$. The 10 brass plates were 10 grooves wide and each plate was skived from CDA 2  brass\footnote{70\% Cu, 30\% Zn, 1/2 hard temper, by Interplex, East Providence, RI.}.
Embedded in this absorber structure were 64 optical fibers, 32 scintillating fibers\footnote{Polystyrene based SCSF-78, produced by Kuraray.} and 32 clear plastic fibers\footnote{PMMA based SK40, produced by Mitsubishi.}. All fibers had an outer diameter of 1.0 mm, the cladding thickness was 20 $\mu$m. Figure \ref{module} shows how these fibers were arranged inside the absorber structure.

The metal absorber thus made up 49\% of the calorimeter volume. The fibers represented 35\% of the instrumented volume (\ie the 12 $\times$ 12 mm$^2$ region in which the fibers were embedded), while air accounted for the remaining 16\%. The effective radiation length ($X_0$) and the Moliere radius ($R_M$)
of the instrumented volume amounted to 29 mm and 31 mm, respectively. The calorimeter was thus $39 X_0$ deep. 

The fibers sampled the electron showers developing in a region with an effective radius of only 6.8 mm, or 0.22 $R_M$. According to GEANT simulations of em shower development in this structure, typically $\sim 45\%$ of the shower energy was deposited in the active volume when an electron entered the calorimeter in its central region
(see Figure \ref{evdisplay}).

\begin{figure}[htbp]
\epsfysize=7.5cm
\centerline{\epsffile{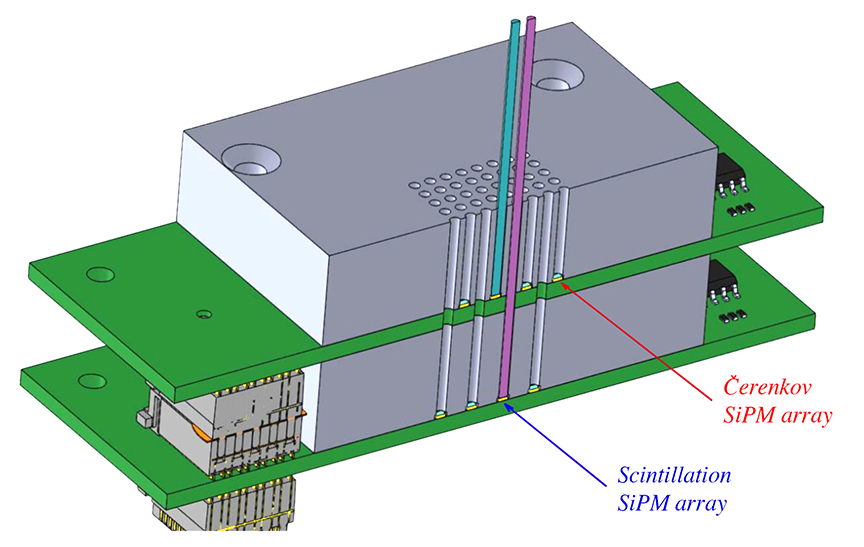}}
\caption{\footnotesize Readout (schematic) of the calorimeter with two arrays of SiPMs. Each fiber is connected to its own SiPM. The location of the SiPM arrays is indicated.}
\label{readout}
\end{figure}
%

Each of the 64 fibers of this calorimeter was interfaced to a single SiPM. The sensors were mounted on a two-tier structure, in a chessboard like arrangement; an exploded view is shown in Figure \ref{readout}. The signals from the 32 \v{C}erenkov fibers were read out by the SiPM mounted on the front tier; through-holes interleaved with the sensors were guiding the scintillating fibers to the corresponding SiPM on the back tier. The boards were equipped with HAMAMATSU S13615-1025 sensors, featuring an active area of 1$\times$1 mm$^2$ and a pitch of 25 $\mu$m, for a total number of 1584 cells/sensor. 
Each SiPM was read out with a simple DC coupled pre-amplifier with a 1 $\mu$s shaping time,  followed by an AC coupled differential amplifier to match the dynamic range of the digitizer. 

\begin{table}[htbp]
\centering
\caption{Main parameters of the SiPM in use. The listed peak sensitivity of 25\% is obtained for the mentioned operating voltage, and a wavelength of 450 nm. 
The breakdown voltage was determined  by measuring the gain-voltage dependence. Even though this is not the unique definition, it is possibly the relevant quantity for the user \cite{Chm17}.}
\vskip 5mm
\begin{tabular}{ |l|c|}
  \hline
   \multicolumn{2}{|c|}{HAMAMATSU S13615-1025} \\
  \hline
 Sensitive area & 1$\times$1 mm$^2$  \\
 Cell pitch & 25 $\mu$m \\
 No. of pixels & 1584 \\
 Peak Photon Detection Efficiency & 25\% \\
 Breakdown voltage V$_{br}$& 53 V \\
 Recommended operational voltage V$_{op}$& V$_{br}$ + 5V \\
 Gain at V$_{op}$ & 7 $\times$ 10$^5$ \\
 Dark Count Rate at V$_{op}$ & 50 kps \\
 After Pulse Rate at  V$_{op}$ & 2 - 3\% \\
 Optical Crosstalk at V$_{op}$ & 1\% \\
   \hline
  \end{tabular}
  \vskip 2mm
    \label{tab:SiPM}
  \end{table}

The ``shadow" of the (circular) fiber tip covered an area of 0.79 mm$^2$, corresponding to 1244 cells. However the SiPMs had a glass front cover of 0.3 mm thickness and the light exited the fiber in a cone defined by its numerical aperture (0.55) and the distance traveled by the light\footnote{The opening angle of this cone depends on the distance traveled by the light in the fiber, for example because of imperfections in the quality of the core/cladding interface and the contribution of cladding light \cite{hartjes}.}. Therefore, the outgoing light was expected to illuminate most of the sensor area, possibly not in a uniform way. 

The SiPMs, in chip size packaging technology, were mounted with a pitch of 1.8 mm. The main features of the sensors are listed in Table \ref{tab:SiPM}.
The two-tier board onto which all 64 SiPMs were soldered also provided individual bias and on-board temperature measurements. 
The used HV generator allowed a fine control for each channel in the 0 - 3V range, so that SiPM response equalization and temperature compensation could be achieved. 

\begin{figure}[b!]
\epsfysize=7.7cm
\centerline{\epsffile{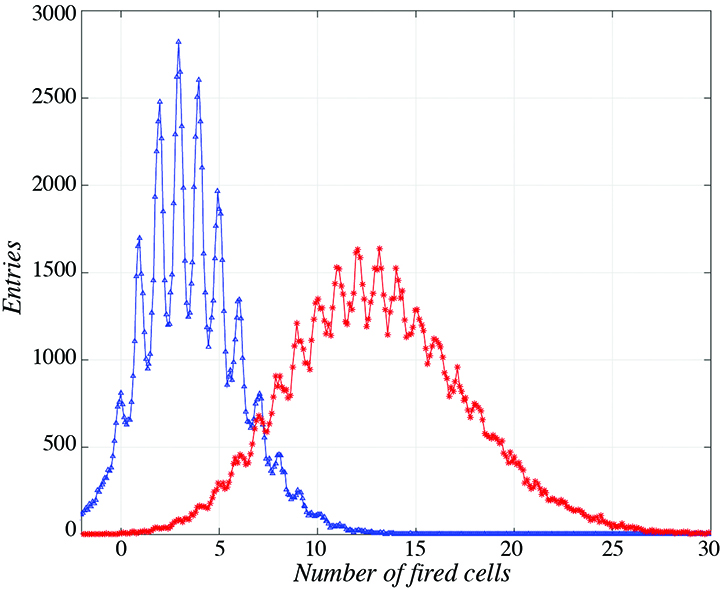}}
\caption{\footnotesize{Signal distributions of the SiPMs in response to light signals used for calibration purposes.}}
\label{SiPMcal}
\end{figure}

The calibration of the SiPMs was greatly simplified by the fact that the signal distributions exhibited a structure that made it possible to count the number of fired cells. This is one of the strong points of SiPMs compared with PMTs.
The SiPMs were calibrated by analyzing the response of each sensor to a large statistics sample of nanosecond long light pulses that conveyed a small number of photons onto the sensitive area. Recording, digitizing and integrating the signals in synchronous mode with respect to the light emission allowed to measure the correspondence between ADC channels and fired cells and study their dependence on the operational voltage. Exemplary spectra for two sensors are shown in Figure \ref{SiPMcal}. The peaks correspond to the different number of fired cells and the shape of the spectrum measures the Poissonian properties of the emitted light, convoluted with detector effects (notably optical cross talk and after pulses) \cite{Vale1,Chmill,Vino,Vale2}. 

Once the parameters of the amplification and DAQ system were known, the peak-to-peak distance could be turned directly into the SiPM gain. 
Moreover, the peak-to-peak value was the gauge to turn the digitized signal into the (raw) number of fired cells. In order to cope with non-linearities, which occur whenever the probability that more than one photon hits an individual SiPM cell is not negligible, the raw estimate of the number of fired cells was corrected throughout the analyses as follows \cite{Sto07,NonLin}:

\begin{equation}
N_{\rm fired}=N_{\rm cells} \times \Bigl[ 1 - \exp\bigl[{-\frac{N_{\rm photons}  \times {\rm PDE}}{N_{\rm cells}}\bigr]}\Bigr]
\label{satur}
\end{equation}

where $N_{\rm fired}$ is the raw number of fired cells, $N_{\rm cells}$ is the number of cells in the sensor, $N_{\rm photons}$ is the actual number of photons in the detected pulse and PDE is the Photon Detection Efficiency at the operational voltage. It should be emphasized that this formula is an approximation, since it applies to ideal circumstances that are not exactly met in practice. For example, it is based on the assumption that all pixels of the SiPM are uniformly irradiated, that all photons arrive simultaneously in time, that there is no prompt cross talk and that there are no effects of pulse recovery. Some aspects of our experimental results, such as the small residual non-linearity (Section 4.2.2) might be a consequence of the inadequacy of the applied corrections.

\begin{figure}[htbp]
\epsfysize=7.3cm
\centerline{\epsffile{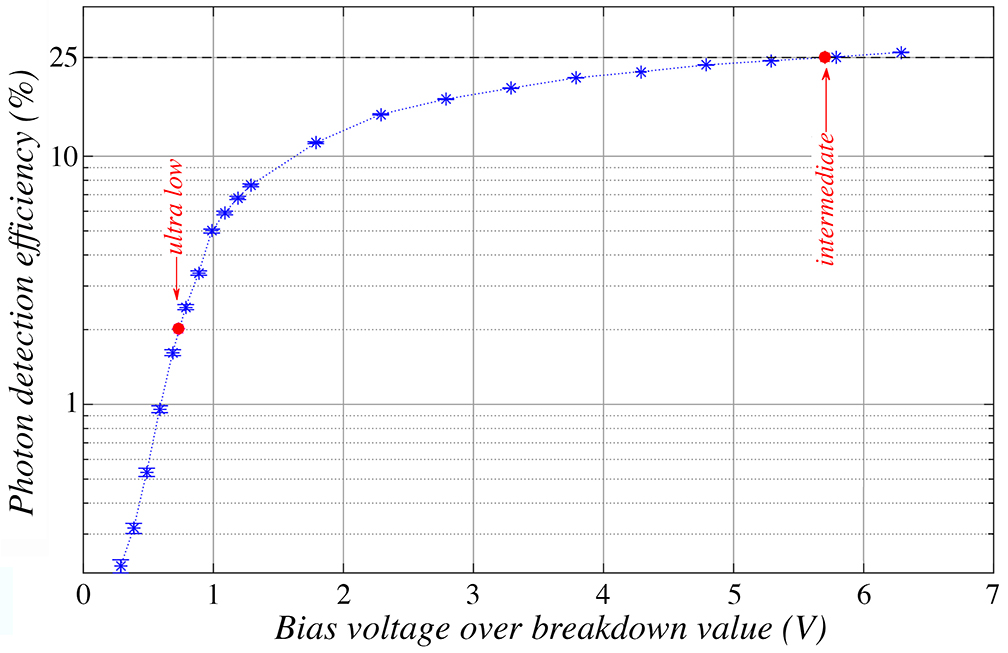}}
\caption{\footnotesize The photon detection efficiency of the SiPM's as a function of the applied bias voltage. The measurements were performed with light with a wavelength of 523 nm. Voltage values at which the beam measurements discussed in this paper were performed are indicated with arrows.}
\label{pde}
\end{figure}

An important aspect for the analyses of the data collected in these experiments was the Photon Detection Efficiency (PDE) of the SiPMs. 
This PDE depended sensitively on the applied bias voltage, as illustrated in Figure \ref{pde}. To limit the effects of saturation, and the resulting signal non-linearity, the SiPMs that detected the scintillation light were operated at a lower voltage ({\sl ultra-low}, see Figure \ref{pde}) than the SiPMs that detected the \v{C}erenkov photons ({\sl intermediate}). 
The gains at these voltages were measured to be $9.9\cdot 10^4$ and $8.0\cdot 10^5$, respectively.
Other factors that contributed to the PDE were the operating temperature and the wavelength of the detected light.
Since it was not possible to operate the two SiPM arrays simultaneously at different bias voltages, the measurements of the \v{C}erenkov and scintillation signals were performed in different runs, in which the bias voltage was optimized for the signals in question.

For the analyses of the experimental data, we needed 
\begin{enumerate}
\item the peak-to-peak distance (see Figure \ref{SiPMcal}) for the conditions used in the various measurements (\ie bias voltage, temperature), and 
\item the ratio of the PDE's at the intermediate and ultra-low bias voltage settings, which were used for the detection of the \v{C}erenkov and scintillation signals, respectively. For the same operating temperature, the ratio of the PDE's was found to be 12.5\footnote{In principle, this ratio is also affected by the different spectra of the two types of light, in combination with the wavelength dependence of the PDE. However, based on the measured characteristics, this effect was found to be negligibly small.}.
\end{enumerate}

The ADC-to-cell conversion can be monitored in real time and adjusted, if necessary, for temperature induced variations of the breakdown voltage and the gain. This can be done either by including in the set-up a light source or by using the spectrum of cells fired by thermally generated charge carriers (dark counts). Showers inducing small signals can also be used. The relevant point here in favor of the SiPM is the possibility to have a calibration, a gain adjustment and a monitor system based on the intrinsic properties of the sensors.

\begin{figure}[htbp]
\epsfysize=5cm
\centerline{\epsffile{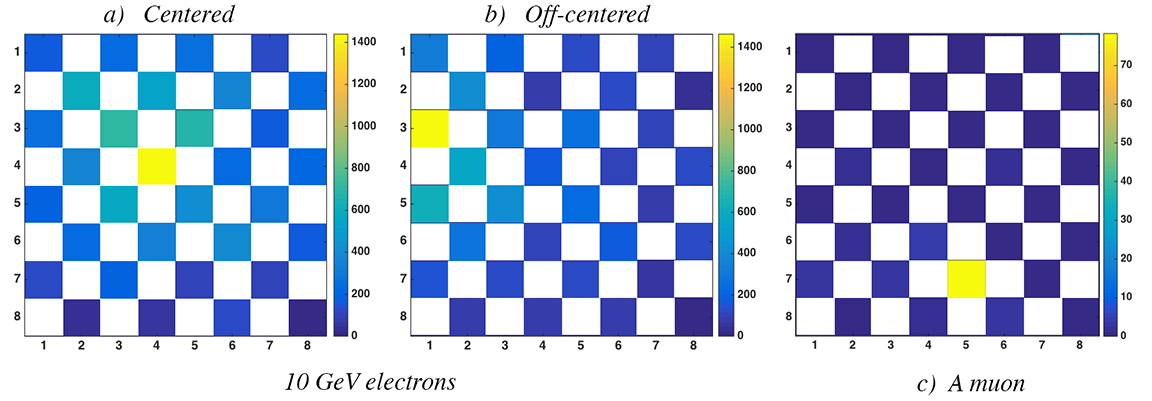}}
\caption{\footnotesize{Event displays in the 8$\times$8 SiPM array for a 10 GeV electron shower and a muon traversing the calorimeter.
The checker board appearance of these event displays reflects the fact that only signals from SiPMs connected to a scintillating fiber are shown.
Each of the white fields is connected to a \v{C}erenkov fiber, of which the signals are not shown here. For color see the online version.}}
\label{events}
\end{figure}

The imaging properties of this module are illustrated in Figure \ref{events}, which shows several event displays for 10 GeV particles, measured with the scintillating fibers. Figure \ref{events}a depicts an electron shower developing in the central region of the calorimeter, while Figure \ref{events}b shows an electron shower located slightly off-center. The event displayed in Figure \ref{events}c concerns a muon traversing the calorimeter. The checker board appearance of these event displays reflects the fact that only signals from SiPMs connected to a scintillating fiber are shown. Each of the white fields was connected to a \v{C}erenkov fiber, of which the signals are not shown in this figure in order to illustrate the energy deposit profiles more clearly.

\section{Experimental setup and measurements}
\vskip-5mm
\subsection{Detectors and beam line}
\vskip -5mm

For these  studies, which were carried out in July 2017, we used secondary or tertiary beams derived from the 400 GeV proton beam delivered by the CERN Super Proton Synchrotron. These particle beams were steered through the H8 line into the dual-readout fiber calorimeter. 

The experimental setup contained, apart from the SiPM calorimeter described in the previous section, a number of auxiliary detectors, which were intended to limit and define the effective size of the beam spot and to determine the identity of individual beam particles. 
Figure \ref{layout} shows a schematic layout of the experimental setup, in which the positions of these auxiliary counters are indicated (not to scale): 

\begin{figure}[htbp]
\epsfysize=3.5cm
\centerline{\epsffile{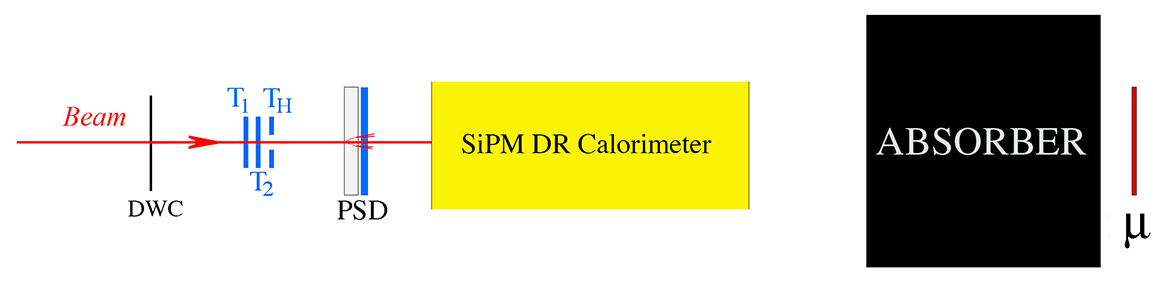}}
\caption{\footnotesize Schematic layout of the experimental setup (not to size). Shown are the delay wire chamber (DWC), the trigger counters (TC), the preshower detector (PSD) and the muon counter ($\mu$). }
\label{layout}
\end{figure}
\begin{itemize}
\item A set of three small scintillation counters provided the signals that were used to trigger the data acquisition systems.
These trigger counters were 2.5 mm thick, the area of overlap between the first two ($T_1,T_2$) was 4$\times$4 cm$^2$. Downstream from these counters, a third scintillation counter ($T_H$) was installed. The latter had a hole with a radius of 10 mm in it. A (anti-)coincidence between the logic signals from these counters provided the trigger (${T_1} \cdot {T_2} \cdot {\overline{T_H}}$).

\item A small delay wire chamber (DWC) made it possible to determine the location of the impact point of the beam particles at the calorimeter surface with a precision of a few mm, depending on the beam energy.

\item About 20 cm upstream of the calorimeter, a preshower detector (PSD) provided signals that could be used to identify the electrons in the beam.
This PSD consisted of a 5 mm thick lead plate, followed by a 5 mm thick plastic scintillator. Electrons started developing showers in this device,
while muons and hadrons typically produced a signal characteristic of a minimum ionizing particle (mip) in the scintillator plate.

\item About 20 m downstream of the calorimeter, behind an additional $8 \lambda_{\rm int}$ worth of absorber, 
a 50$\times$50 cm$^2$ scintillation counter ($\mu$) served to identify the muons in the particle
beams. 
\end{itemize}

\subsection{Data acquisition}
\vskip -5mm
In order to minimize delays in the DAQ system, short, fast cables  were used to transport the signals from the trigger counters to the counting room. All other signals were transported through cables with (for timing purposes) appropriate lengths.

We used two independent different data acquisition systems for these measurements, one system dealt with the SiPM data, another with the signals from
the auxillary detectors. Both systems used the same trigger (${T_1} \cdot {T_2} \cdot {\overline{T_H}}$). Offline, the information from both systems was synchronized and merged into ntuples that were used for the data analyses. 

In the counting room, signals from the PSD and the muon counter were integrated and digitized with a sensitivity of 100 fC/count and a 12-bit dynamic range  in a charge ADC (CAEN V862AC).
The signals from the wire chamber were recorded with 140 ps resolution in a 16-channel  CAEN V775N TDC, and converted into ($x,y$) coordinates of the point where the beam particle traversed the chamber.  
This data acquisition system used VME electronics.
All information was collected using gate widths of the order of 100 ns, and read out event-by-event through the V2718 CAEN optical link bridge with a dead time of $\sim 300~\mu$s. The event rates were such that pileup effects were negligible.

\begin{figure}[htbp]
\epsfysize=7.5cm
\centerline{\epsffile{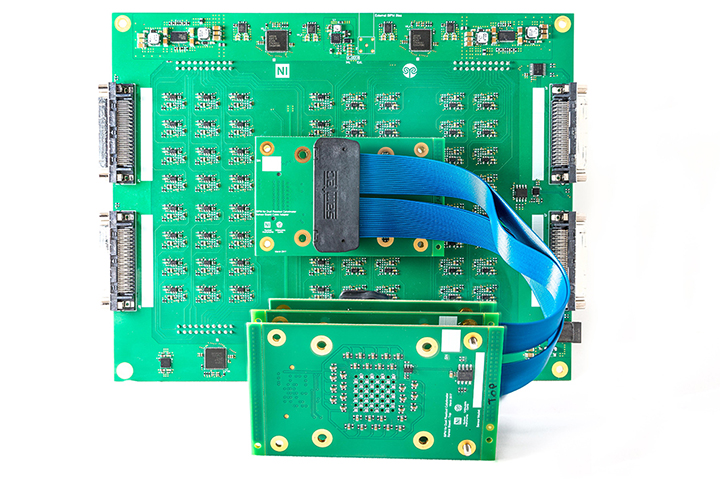}}
\caption{\footnotesize The front-end electronics of the SiPM detectors. In the foreground,  the two-tier boards hosting the sensors are shown. These are interfaced with the fibers, as shown in Figure \ref{readout}. In the background, the mother board is shown. This contains the amplifiers, the shapers and the sensor bias generator/control.}
\label{FEE}
\end{figure}

The signals from the two-tier board were fed into a mother board through a 64-channel coaxial cable with an adapter board (Figure \ref{FEE}). The mother board hosted 64 DC-coupled amplifiers with a $1 \mu$s shaping time, compliant with the expected event rate. 
The channels were read out using a Multichannel Analog to Digital Acquisition System (MADA) \cite{mada_ref}. Each of two boards digitized 32 channels at a rate of 80 MS/s  and 14-bit ADCs performed real-time charge integration on FPGAs.

Our readout scheme optimized the CPU utilization and the data taking efficiency using the bunch structure of the SPS accelerator cycle (which lasted between 36 and 54 s, depending on the various tasks of the accelerator complex), during which period beam particles were provided to our experiment by means of either one or two extractions with a duration of 4.8 seconds each.

\subsection{Experimental data, calibration and analysis methods}
\vskip-5mm

For the measurements described in this paper, we used either 60 or 180 GeV secondaries, produced by 400 GeV protons from the SPS accelerator on a target shared by several beam lines. Low energy tertiary
beams were produced off a target installed in the 60 GeV secondary beam, and beams with energies above 60 GeV were derived similarly from the 
180 GeV secondary beam. The secondary beams were also used to provide intense beams of $\mu^+$ particles (obtained by blocking all other particles with upstream absorbers). The tertiary beams had a mixed composition. For energies below 50 GeV, they consisted primarily of electrons, with small admixtures of hadrons and muons. For higher energies, pions gradually became a very significant contribution and at energies above 100 GeV, muons were dominant.
As described below, dedicated efforts were made to extract pure electron and muon event samples from the collected data.

\begin{figure}[htbp]
\epsfysize=10cm
\centerline{\epsffile{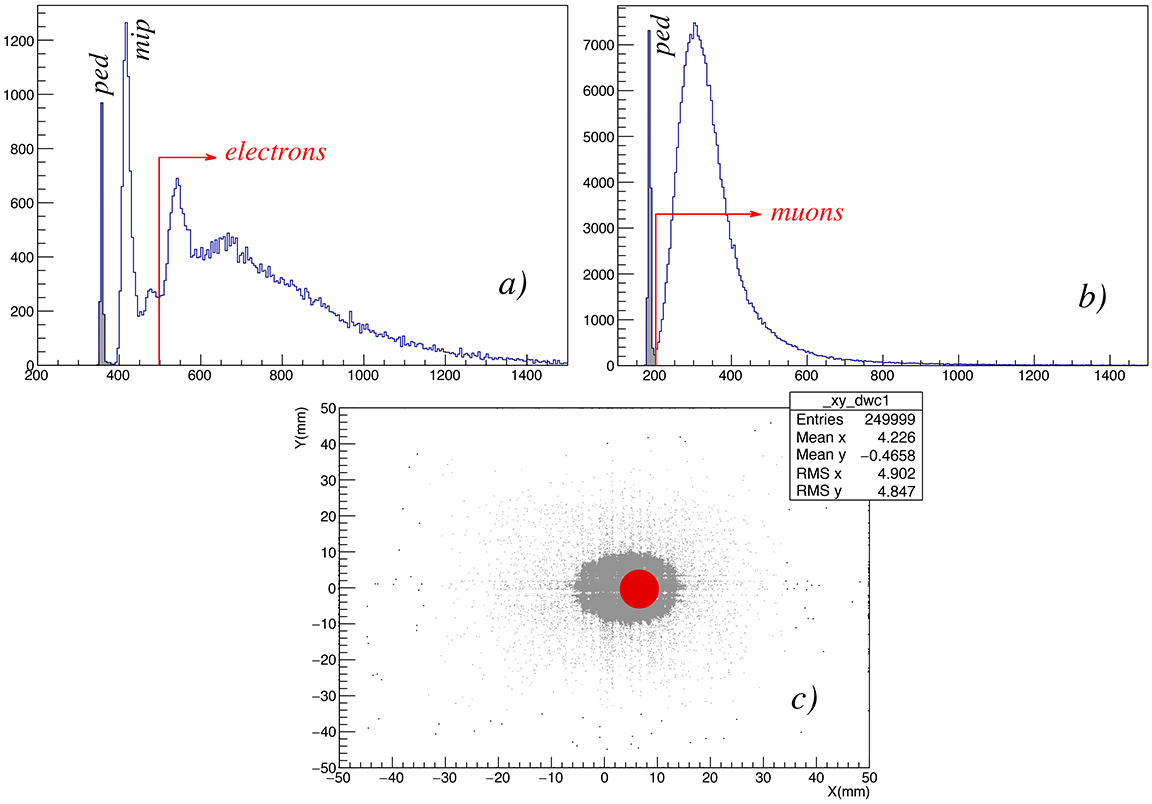}}
\caption{\footnotesize The signal distributions measured in the preshower detector ($a$) or muon counter ($b$) were used to select pure samples
of electrons or muons, respectively. The data from the delay wire chamber ($c$) could be used to select events that entered the calorimeter in its central region, represented by the red spot.}
\label{psdmu}
\end{figure}

Dedicated runs with tertiary beams were carried out for the following energies: 6, 10, 20, 30, 40 and 50 GeV (from 60 GeV secondaries) and 40, 60, 80, 100 and 125 GeV (from 180 GeV secondaries).
Off-line, the beam chamber information could be used to select events within a small beam spot covering the central region of the calorimeter (typically with a radius $< 3$~mm, see Figure \ref{psdmu}c). Alternatively, the energy deposit pattern in the calorimeter itself could be used for this purpose, for example by selecting events in which the fiber with the largest signal was located in the central $4\times$4 fiber region (see Figure \ref{module}).
The information provided by the other auxiliary detectors was used to identify and select the desired particles.

Muon event samples were selected on the basis of the signals in the muon counter. Figure \ref{psdmu}b shows the signal distribution in this counter for
125 GeV beam particles. Events with a signal above the indicated cutoff value were selected for the muon sample.
Electrons were identified as particles that produced a signal in the PSD that was larger than $\sim 150$ ADC counts above pedestal, equivalent to the combined signals from at least three minimum ionizing particles (mips) traversing this detector. Figure \ref{psdmu}a shows a typical signal distribution (for 10 GeV electrons) in the PSD, with the mentioned cutoff value.
An additional requirement for electron events was that no signal incompatible with electronic noise (\ie the pedestal) was produced in the muon counter. 

\section{Experimental results}
\vskip -5mm
\subsection{Crosstalk}
\vskip -5mm
Because of the large difference in light yield between the \v{C}erenkov and scintillation fibers, crosstalk is a major concern.
For example, if this light yield difference is a factor of 50, then the \v{C}erenkov signals would increase by 50\% if 1\% of the scintillation light was detected by the SiPMs that read out the signals from the \v{C}erenkov fibers.
In order to find out if there is a contribution of scintillation light to the \v{C}erenkov signals, the best way would be a measurement
in which the scintillating fibers are physically removed and compare the results with the \v{C}erenkov signals measured in the setup described
in Section 2. Since that was not possible without completely rebuilding the calorimeter, alternative methods were used.

\begin{figure}[htbp]
\epsfysize=7cm
\centerline{\epsffile{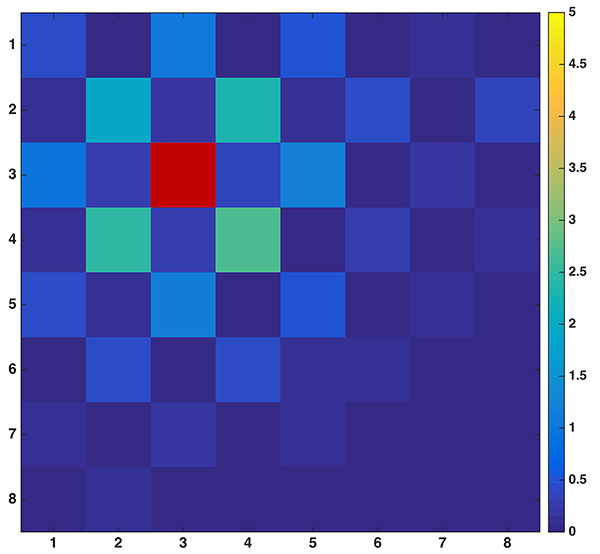}}
\caption{\footnotesize Signal distribution in the 64 sensors, resulting from the illumination of one scintillating fiber with a large light pulse ($\sim 1400$ fired cells). The colors indicate the number of fired cells in the different sensors. The illuminated fiber is indicated in red (for color see the online version). This distribution represents the average 
of about 100,000 events in which the red fiber was illuminated.}
\label{background}
\end{figure}

The first method was performed in the lab, before the module was transported to CERN. 
The crosstalk between scintillating and \v{C}erenkov fibers was studied as follows. At the front face of the calorimeter module, all the tips of the fibers, with the exception of one, were masked. A pulsed LED illuminated the uncovered tip and signals were simultaneously recorded for all 64 sensors.
Figure \ref{background} shows an example of the results obtained for this type of measurement. The non-zero signals are clearly concentrated in the immediate vicinity of the illuminated fiber, for both types of fibers. This particular display represents the average result of about 100,000 events. Similar results were obtained when other fibers were illuminated instead.
Analysis of these data showed that 
when a scintillating fiber was illuminated, the distribution of the sum of the fired cells in the 32 \v{C}erenkov fibers had a mean value of 0.3\% of the scintillation signal, and a rms value of 0.1\%.
Strictly speaking, this observation represents an upper limit to the crosstalk, since it cannot be excluded that a very small fraction of the LED light directly entered 
a neighboring \v{C}erenkov fiber. 
\vskip 2mm

\begin{figure}[htbp]
\epsfysize=7cm
\centerline{\epsffile{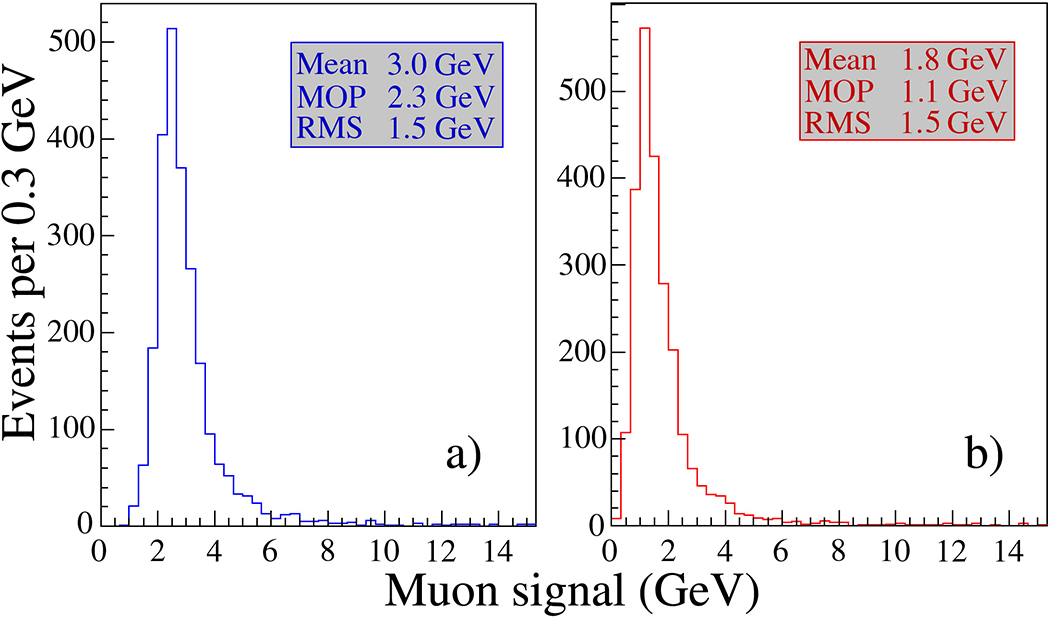}}
\caption{\footnotesize Scintillation ($a$) and \v{C}erenkov ($b$) signal distributions for 100 GeV $\mu^+$ measured in the (copper based dual-fiber) DREAM calorimeter with PMT readout. The energy scale was determined with electrons, separately for the scintillation and the Cherenkov signals \cite{DREAMmu}.}
\label{mupmt}
\end{figure}

The second method was carried out in the beam line at CERN.
This method was based on the fact that the signals from muons are quite different for the two types of fibers. This is illustrated by Figure \ref{mupmt}, which shows the signal distributions for 100 GeV $\mu^+$, measured with a calorimeter module of comparable composition, read out with PMTs. This module was calibrated with electrons, and the response to these particles, \ie the average signal per unit deposited energy, was equalized for the scintillation and 
\v{C}erenkov signals. The horizontal scale in Figure \ref{mupmt} is based on this energy calibration.

The figure shows substantial differences between the response functions for the two signals from muons. Both the average signal and the most probable (mop) signal are 1.2 GeV smaller for the \v{C}erenkov signals. The DREAM Collaboration found that this difference was constant for muons of different energies, ranging from 40 - 200 GeV \cite{DREAMmu}.
The explanation of this phenomenon is the fact that the \v{C}erenkov fibers do not produce a signal for the ionization part of the energy loss of the muons in the calorimeter, since the \v{C}erenkov light produced by the muons falls outside the numerical aperture of the fibers.
The \v{C}erenkov fibers are only sensitive to the radiative processes (bremsstrahlung) that also contribute to the energy loss, and there is no reason why this component of the signals should be different for the two types of fibers.
The energy lost by minimum ionizing particles in the DREAM calorimeter structure was estimated to be $\sim$7 MeV/cm, \ie 1.4 GeV for the total, 2 m long detector. This was found to be in good agreement with the difference measured between the signals from the scintillating and \v{C}erenkov fibers, both for what concerns the average and the most probable signal values.

This phenomenon also offered a possibility to measure crosstalk effects in the SiPM calorimeter module. 
Based on the results obtained with PMT readout, one would expect for 125 GeV $\mu^+$ to find the most probable energy deposit measured with the \v{C}erenkov fibers in the SiPM calorimeter to be a factor 2.03 (\ie 2.271/1.117) smaller than that measured with the scintillating fibers.
Given the length of the detector ($39 X_0$ \vs $100 X_0$) and the measured effects in the DREAM calorimeter, one would also expect the difference between the scintillation and \v{C}erenkov signals to be $\sim 0.45$ GeV.
Any larger signal for the \v{C}erenkov fibers, or any smaller difference with the scintillation signals, would indicate a contribution of crosstalk to the
\v{C}erenkov signals.

\begin{figure}[htbp]
\epsfysize=7cm
\centerline{\epsffile{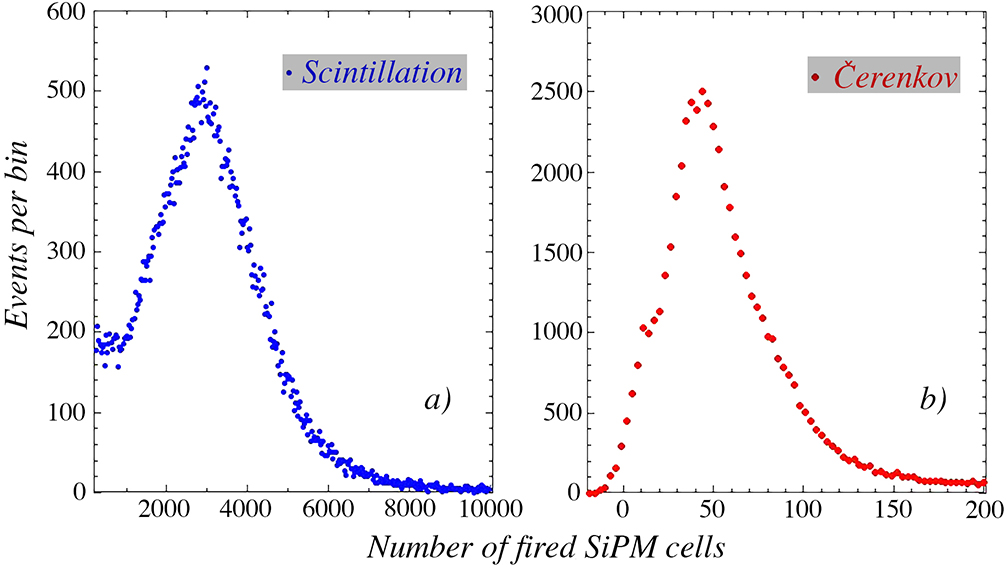}}
\caption{\footnotesize Scintillation ($a$) and \v{C}erenkov ($b$) signal distributions for 125 GeV $\mu^+$ measured in the brass based dual-fiber calorimeter with SiPM readout.}
\label{muSiPM}
\end{figure}

Figure \ref{muSiPM} shows the signal distributions for 125 GeV $\mu^+$ measured with the SiPM calorimeter. The scintillation signals were corrected for the effects of signal saturation. After this correction, the most probable muon signal consisted of 2,960 fired cells (Figure \ref{muSiPM}a). Using a measured scintillation light yield of 3,200$\pm$200 photoelectrons per GeV deposited energy (see Section 4.2.2), this corresponds to a most probable energy deposit of 2,960/3,200 = 0.93$\pm$0.06 GeV. According to the above considerations, one would expect, in the absence of crosstalk and given the measured scintillation signal, a most probable \v{C}erenkov signal equivalent to an energy deposit of 0.47$\pm$0.04 GeV (namely, 0.46$\pm$0.03 GeV based on the $S/C$ signal ratio, 0.48$\pm$0.06 GeV based on the $S - C$ value).

Saturation did not play a role at all for the \v{C}erenkov signals. The most probable signal was observed to consist of 44 fired cells. After correcting for the contribution of thermal noise (\ie signals observed in the absence of light, on average corresponding to 6 fired SiPM cells), we concluded that the most probable \v{C}erenkov signal produced by the muons traversing the calorimeter consisted of 38 fired cells (Figure \ref{muSiPM}b). 

It turned out to be non-trivial to translate this signal into an energy deposit that may be compared to the expected value (0.47$\pm$0.04 GeV), since this depends on the \v{C}erenkov light yield used as the basis for the conversion. This light yield is based on the \v{C}erenkov signals measured for electron showers in the calorimeter module, 28.6 Cpe (\v{C}erenkov photoelectrons) times the energy of the beam particle (Section 4.2.1). Given that 45\% of the shower energy is deposited in the module, the muon signal thus corresponds to an energy deposit of $(38/28.6)\times 0.45 = 0.59$ GeV.
However, as discussed in the Appendix, simulations showed that only 36\% of the total \v{C}erenkov light generated by an em shower in an infinitely 
large absorber would be generated in the area covered by our small module. If we took that light yield as the basis for the conversion, then the muon signal would  correspond to an energy deposit of $(38/28.6)\times 0.36 = 0.48$ GeV. The question is thus if the muon signals measured in the small module represent only a small fraction of the total signal that would have been observed if the calorimeter had been much larger, or if enlarging the calorimeter (laterally) would have made no difference for the muon signals\footnote{As shown in the Appendix, the answer to this question is inconsequential for the energy deposit derived from the scintillation signals.}. In the latter case, the measured energy deposit (0.48 GeV) would be compatible with {\sl no} crosstalk, whereas in the first case $\sim 0.4\%$ of the scintillation light (corresponding to 12 photoelectrons) would have contributed to the measured \v{C}erenkov signals.

This intrinsic uncertainty associated with the interpretation of these results led us to the conclusion that they are not incompatible with those obtained in the lab tests, and we have chosen the latter (0.3$\pm$0.1\%) for determining the crosstalk level.

\subsection{Light yield}
\vskip -5mm
Thanks to the fact that this calorimeter offers the possibility to count the fired cells, calibration of the signals from em showers was straightforward. However, in order to determine the light yield, \ie the absolute response in photoelectrons per GeV deposited energy, it is important to determine what fraction of the showers was deposited in the small calorimeter. We used GEANT4 simulations \cite{geant4} for that purpose\footnote{Version GEANT4.10.3.p01, with physics list FTFP\_BERT}.

\subsubsection{The \v{C}erenkov signals}

\begin{figure}[htbp]
\epsfysize=7.8cm
\centerline{\epsffile{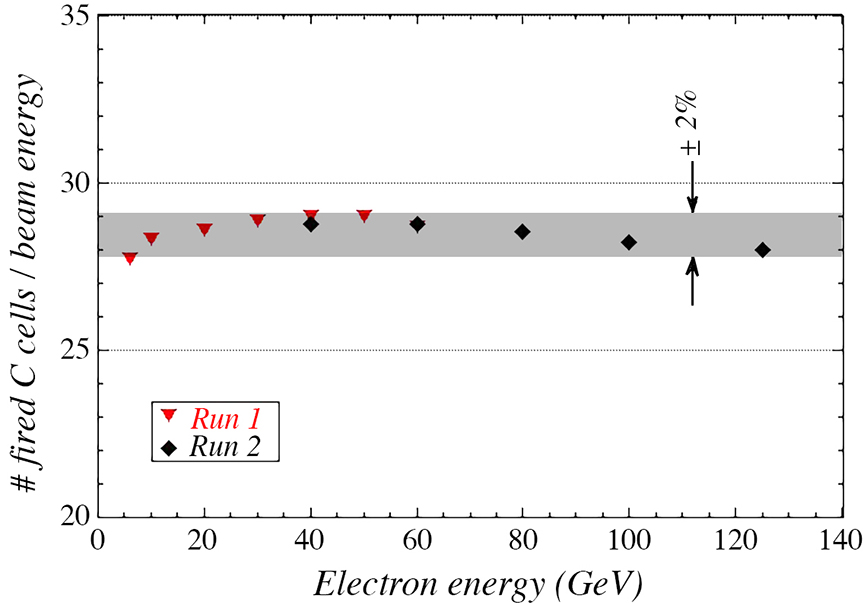}}
\caption{\footnotesize{The average number of \v{C}erenkov photoelectrons measured in the SiPMs divided by the electron beam energy (Cpe/GeV), as a function of the electron energy. These results were obtained with a bias voltage of 5.7 V above the breakdown value (Figure \ref{pde}). The shaded area represents deviations of less than 2\% from the average value.}}
\label{Clin}
\end{figure}

\begin{figure}[htbp]
\epsfysize=8cm
\centerline{\epsffile{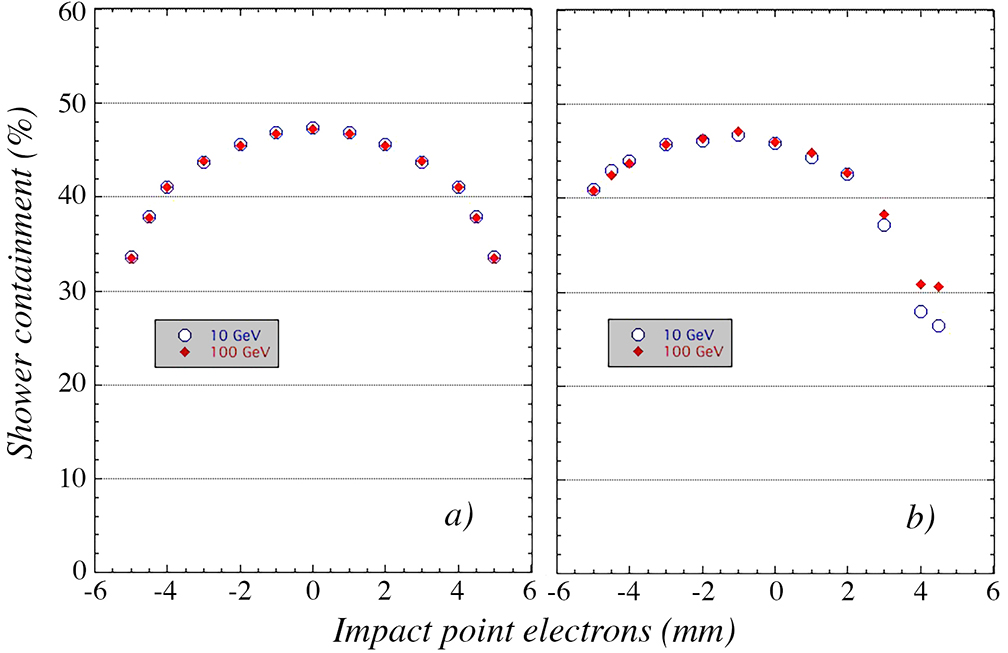}}
\caption{\footnotesize{The average fraction of the shower energy deposited in the SiPM calorimeter, as a function of the impact point of the
10 GeV and 100 GeV electrons used for these GEANT4 simulations. Results are given for electrons that entered the calorimeter along the direction of the fibers ($a$) or at an angle of 0.2$^\circ$ in both the horizontal and the vertical plane ($b$).}}
\label{simul}
\end{figure}

Figure \ref{Clin} shows the average number of detected photoelectrons as a function of the electron beam energy, divided by that energy, for the \v{C}erenkov channel. This number was measured to be approximately constant, at $\sim$28.6 Cpe/GeV, over the entire range of 6 - 125 GeV for which measurements were performed, with a standard deviation of 0.4 Cpe/GeV. This indicates two things:
\begin{enumerate}
\item There was no saturation in the \v{C}erenkov signals
\item The average shower containment was independent of the electron energy
\end{enumerate}
In principle it is possible that deviations from these two conditions conspired to yield the measured result. However, since the average shower containment was found to be independent of the electron energy (Figure \ref{simul}a), we conclude that the calorimeter was linear to within 2\% for what concerns the \v{C}erenkov signals, over the full energy range at which it was tested.

Results of the simulations are shown in Figure \ref{simul}, for 10 GeV and 100 GeV electrons. The average fraction of the electron energy deposited in the SiPM calorimeter is shown as a function of the impact point of the particles. 
Two sets of simulations were performed. In the first set (Figure \ref{simul}a), the electrons entered the detector along the direction of the fibers.
To avoid ``channeling'' effects, in which a beam particle can travel over a very long distance inside an individual fiber, the detector was also rotated over a small angle (0.2$^\circ$ in both the vertical and horizontal plane). The results of this second set of simulations are shown in Figure \ref{simul}b.
Because of the incomplete lateral containment of the showers, the effects of this rotation are clearly visible. However, if the impact points are limited to a region with a radius of about 3 mm around the geometrical center of the calorimeter, the 
containment fraction is rather insensitive to the impact point of the electrons, and the average shower containment is about 45\%, for both sets of simulated data.
The fact that the results shown in Figure \ref{simul}a are essentially identical for 10 and 100 GeV support the statement that the lateral shower containment is 
energy independent. The small differences observed when the module was tilted (Figure \ref{simul}b) reflect the difference in the {\sl longitudinal} shower development.

\begin{figure}[htbp]
\epsfysize=7cm
\centerline{\epsffile{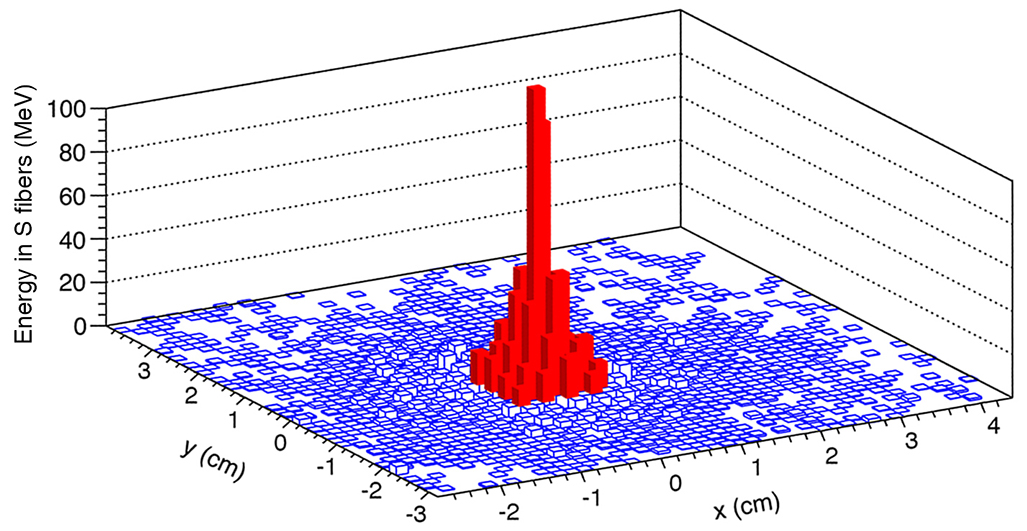}}
\caption{\footnotesize{Event display of a simulated 50 GeV electron shower developing in an extended calorimeter structure of the type used in these tests. The figure shows the signals in the individual scintillating fibers. The white patches represent fibers in which no (measurable) energy was deposited.The fibers in the area covered by our detector are  indicated in red. The vertical scale represents the energy deposited in the individual scintillating fibers. For this event, the total energy deposited in these fibers amounts to 1.982 GeV, of which 0.902 GeV (46\%) is distributed among the red fibers. For color see the online version.}}
\label{evdisplay}
\end{figure}

Figure \ref{evdisplay} shows an event display of a simulated 50 GeV electron shower developing in an extended calorimeter structure of the type used in our tests. 
The lego plot represents the signals in individual scintillating fibers, with the red ones located in the area covered by our small calorimeter. These red fibers detected 46\% of the total scintillation signals produced in this event.

The containment results mean that the intensity of the em shower signals measured with the \v{C}erenkov fibers corresponded to 64$\pm$2  photoelectrons per GeV deposited energy. At face value, this seems more than two times larger than measured for a similar calorimeter with PMT readout \cite{RD52_em}. On the other hand, we should realize that
some fraction of this light is the result of crosstalk. After correcting for this effect, we found a \v{C}erenkov light yield of 54$\pm$5 photoelectrons per GeV deposited energy (see Section 4.2.2). 

This increased light yield should improve the stochastic term in the em energy resolution from 13.9\%/$\sqrt{E}$ to 12.5\%/$\sqrt{E}$, bringing this resolution somewhat closer to the limit set by sampling fluctuations alone (8.9\%/$\sqrt{E}$ \cite{RD52_em}).

\subsubsection{The scintillation signals}

Whereas signal saturation and non-linearity did not play a significant role for the \v{C}erenkov light, this was most definitely different for the 
scintillation signals. These effects turned out to be already noticeable at the low end of the electron energy range studied here, even when the PDE was lowered to only a few percent, by means of the bias voltage (the ultra-low setting, Figure \ref{pde}).

\begin{figure}[htbp]
\epsfysize=8cm
\centerline{\epsffile{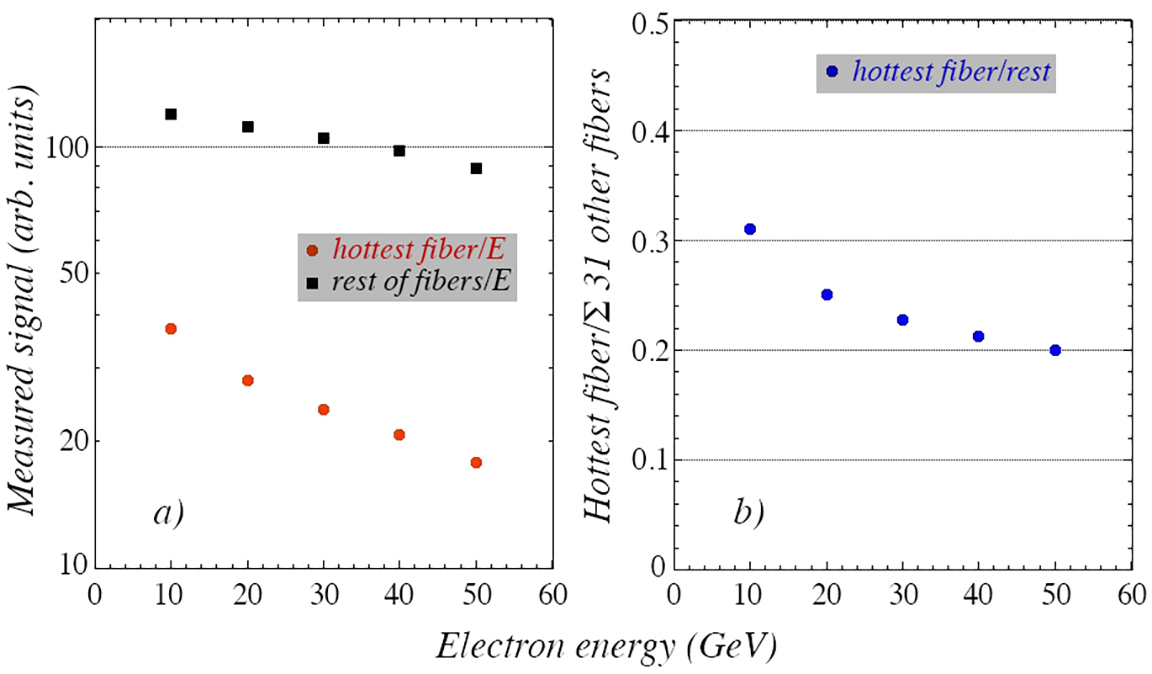}}
\caption{\footnotesize{The average calorimeter signal from the scintillating fibers per unit deposited energy, as a function of the electron beam energy. The quantum efficiency was set very low for these measurements (2\%, see Figure \ref{pde}). Results are shown separately for the hottest fiber and for the sum of the signals measured by the other 31 scintillating fibers ($a$). The ratio of these two signals, as a function of the electron beam energy ($b$).}}
\label{SiSaturation}
\end{figure}

This is illustrated in Figure \ref{SiSaturation}a, which shows the raw data obtained with the electron beam. The average signal divided by the beam energy, which was measured to be constant over the entire energy range from 6 - 125 GeV for the \v{C}erenkov signals, decreased by more than a factor of two in the ``hottest'' fiber, \ie the fiber that measured the largest signals, between electron energies of 10 and 50 GeV. In the rest of the fibers, which recorded much smaller signals, a decrease of 25\% was measured over this same energy range. Figure \ref{SiSaturation}b shows the ratio of these two signals, \ie a measure for the relative contribution of the central fiber to the total signal. Since the lateral shower profile is independent of the electron energy, the decrease of this ratio indicates that this energy dependence was indeed caused by saturation effects. 
\begin{figure}[htbp]
\epsfysize=8cm
\centerline{\epsffile{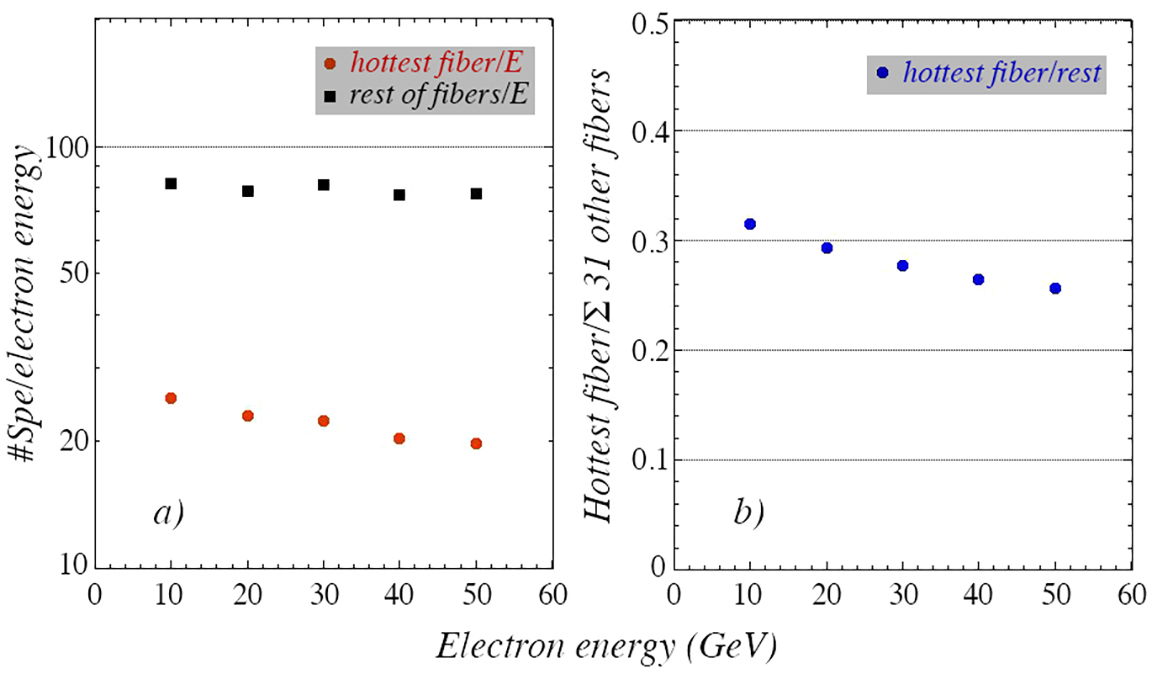}}
\caption{\footnotesize{Number of photoelectrons divided by the electron beam energy, as a function of energy, for the signals from the scintillating fibers (Spe/GeV). 
The signals were corrected for saturation effects with Eq. \ref{satur}.  Results are shown separately for the hottest fiber and for the sum of the signals measured by the other 31 scintillating fibers ($a$). The ratio of these two signals, as a function of the electron beam energy ($b$).}}
\label{Scilight}
\end{figure}

After the signals were corrected for saturation effects, using Eq. \ref{satur}\footnote{The results shown in Figure \ref{Scilight}a were obtained with $N_{\rm cells} = 1584$. We also performed the saturation corrections with $N_{\rm cells} = 1244$ (see Section 2), which led to a slight improvement of the linearity.}, much of the non-linearity disappeared (Figure \ref{Scilight}). The average signals, divided by the electron beam energy, are shown as a function of energy in Figure \ref{Scilight}a, both for the hottest fiber and for the sum of the signals measured by the other 31 scintillating fibers. The signals shown in this figure were also converted into photoelectrons, using the proper calibration for this bias voltage setting. 
Comparing Figures \ref{Scilight}a and \ref{SiSaturation}a, it seems that the non-linearity has indeed more or less completely disappeared in the sum of 31, but that the hottest fiber still exhibits remnants of this effect. This is also evident from Figure \ref{Scilight}b, which shows that the ratio of the signals in the hottest fiber and the 31 neighbors still decreases with the beam energy, albeit to a lesser extent than in Figure \ref{SiSaturation}b. In order to translate these results into a scintillation light yield that may be compared with that measured for the \v{C}erenkov signals, the numbers from Figure \ref{Scilight}a have to be corrected for the actual deposited energy fraction, and also for the (factor 12.5) difference between the PDE values at which the measurements of the two types of signals were carried out.
If one takes the sum of all 32 signals measured for the lowest energy (10 GeV) as the basis for this calculation, then the fraction of the total electron energy contained in the instrumented calorimeter volume (45\%) has to be used. This leads to a light yield of $108 \times 12.5/0.45 \approx 3,000$ photoelectrons per GeV deposited energy. If one uses only the (sum of 31) signals where
non-linearity effects seem to be absent, then the fact that 29\% of the shower energy was deposited in the area covered by these fibers leads to a light yield
of $80 \times 12.5/0.29 \approx 3,400$ photoelectrons per GeV. 
These numbers have not been corrected for possible (minor) contributions due to Geiger discharge in neighboring pixels.

We conclude from these results that the scintillation light yield at a bias voltage of 5.7 V above breakdown was 3,200 $\pm$ 200 photoelectrons per GeV deposited energy. That is thus about 50 times larger than the measured intensity of the signals in the SiPMs connected to the \v{C}erenkov fibers.
As discussed in Section 4.1, the contribution of optical crosstalk to the \v{C}erenkov signals was $\sim 0.3\%$ of the scintillation signals. Using the value obtained in the lab 
measurements (0.3$\pm$0.1\%), this corresponds to 10$\pm$4 photoelectrons. 
After eliminating the effect of crosstalk, we find thus that the \v{C}erenkov light yield of our calorimeter amounted to 54$\pm$5 Cpe/GeV.
This means that $\sim 15\%$ of the measured \v{C}erenkov signals was in fact the result of scintillation light. 
The $S/C$ ratio of our calorimeter is (3,200$\pm$200)/(54$\pm$5) = 59$\pm$8.


\subsection{Shower profiles}
\vskip -5mm

The fact that each fiber was read out separately in the tested calorimeter made it possible to measure the lateral profiles of em showers in unprecedented detail, very close to the shower axis. These profiles were measured as follows. Using the PSD and the muon counter, a clean sample of electrons was selected. For each event, the coordinates of the impact point were determined from the center of gravity ($\bar{x},\bar{y}$) of the energies $E_i$
deposited in the 32 fibers (with position coordinates $x_i,y_i$) that contributed to the total signal:

\begin{equation}
\bar{x}~=~{{\sum_i{x_i E_i}}\over{\sum_i{E_i}}} ,~~~~~\bar{y}~=~{{\sum_i{y_i E_i}}\over{\sum_i{E_i}}}
\label{pos1}
\end{equation}

The radial distance ($r_i$) between each individual fiber $i$ and the shower axis was then determined as 
\vskip-5mm
\begin{equation}
r_i~=~\sqrt{(x_i - \bar{x})^2 + (y_i - \bar{y})^2}
\end{equation}

In this way, the average signal in an individual fiber could be determined as a function of $r$, and this represents the lateral shower profile.
This exercise was performed separately for the two types of signals. For the \v{C}erenkov signals, 40 GeV electrons were used.
In order to limit the effects of signal saturation as much as possible, we chose 10 GeV electrons measured with the ultra-low bias voltage
for the scintillation signals. We want to point out again that the lateral profiles of high-energy electron showers are independent of the 
electron energy.

\begin{figure}[b!]
\epsfysize=11cm
\centerline{\epsffile{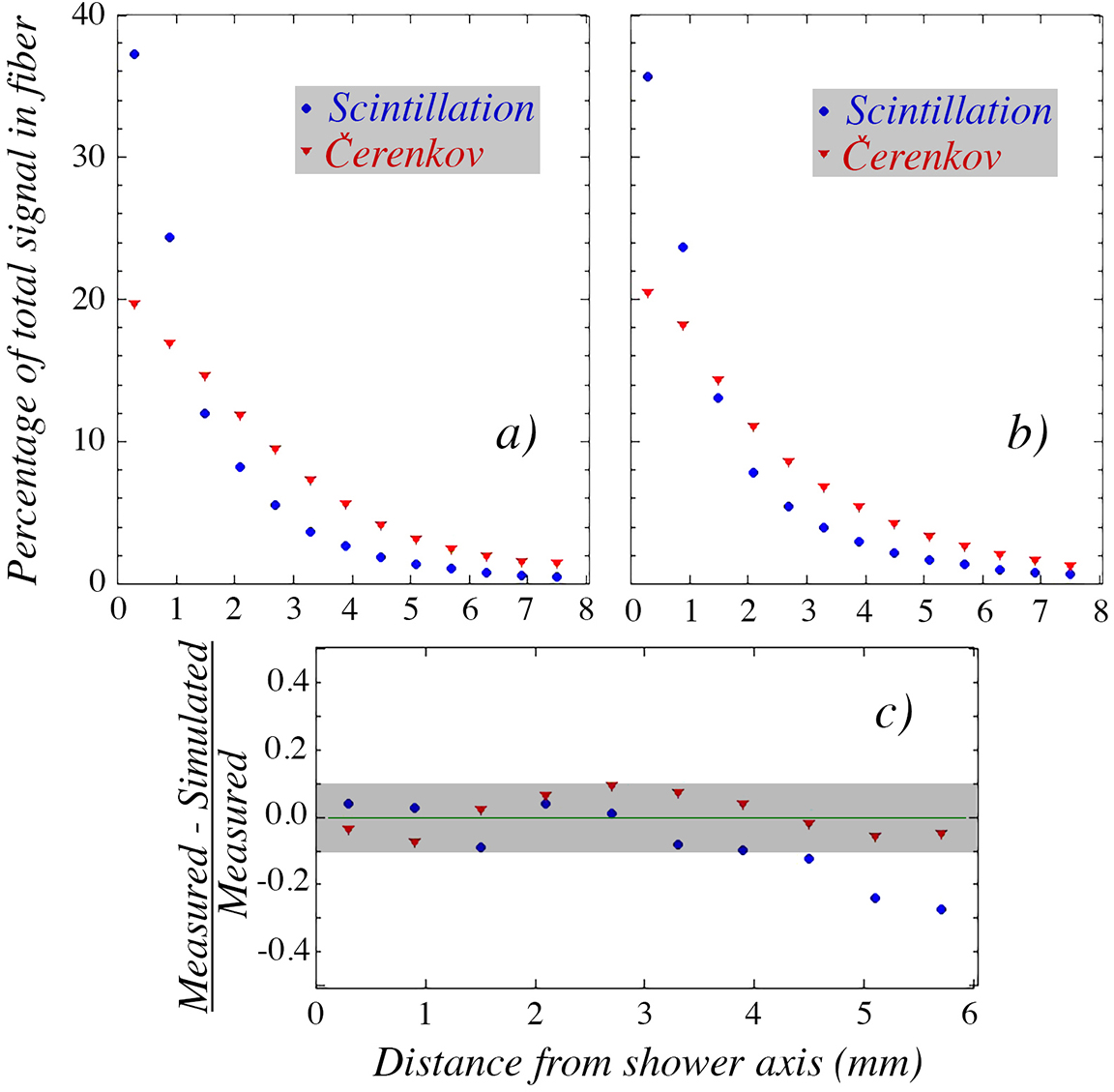}}
\caption{\footnotesize{Lateral profiles of electromagnetic showers in the  brass-fiber dual-readout calorimeter, measured separately with the \v{C}erenkov
and the scintillation signals ($a$). The same lateral profiles simulated with GEANT4 ($b$). Fractional differences between the measured and simulated profiles ($c$). }}
\label{emlprofiles}
\end{figure}

\begin{figure}[htbp]
\epsfysize=9cm
\centerline{\epsffile{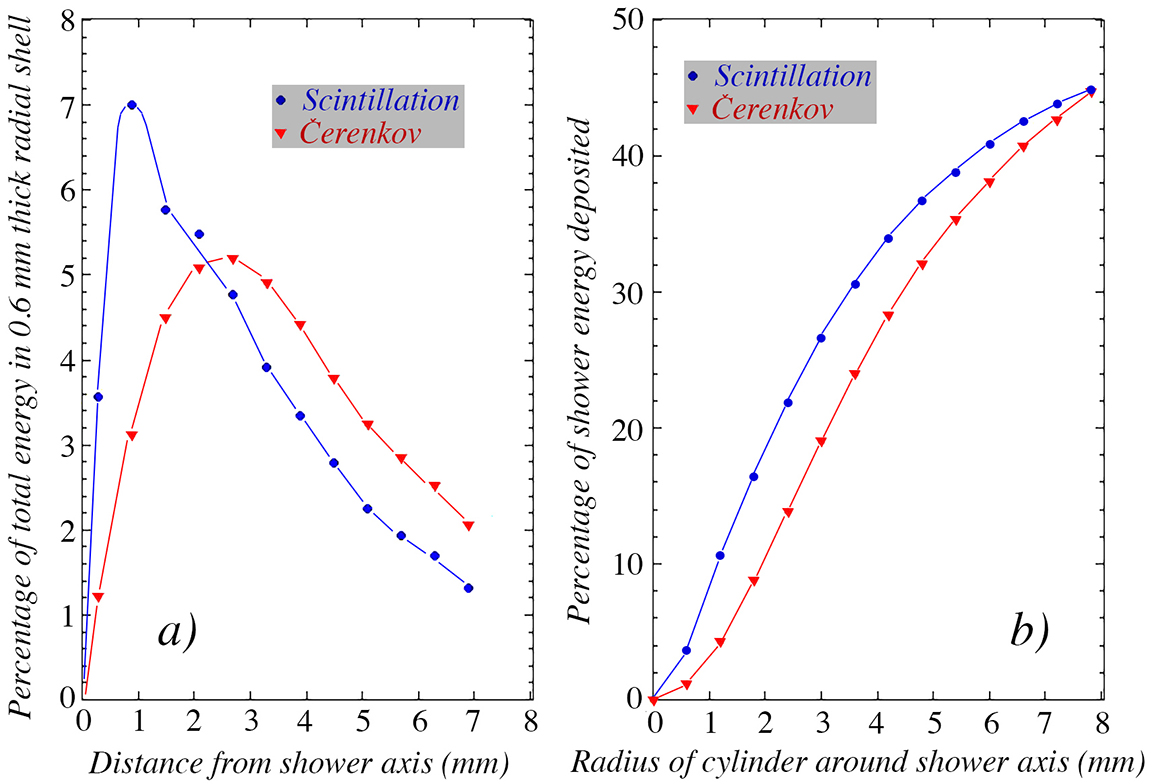}}
\caption{\footnotesize{Radial profiles of electromagnetic showers in the brass-fiber dual-readout calorimeter, measured separately with the \v{C}erenkov
and the scintillation signals ($a$). The fraction of the shower energy deposited in a cylinder around the shower axis as a function of the radius of that cylinder, measured separately with the \v{C}erenkov and the scintillation signals ($b$). The lines are drawn to guide the eye.}}
\label{emrprofiles}
\end{figure}

The profiles are shown in Figures \ref{emlprofiles} and \ref{emrprofiles}.
In Figure \ref{emlprofiles}, the average signal measured in individual fibers is plotted as a function of $r$, \ie  the distance to the shower axis. We call this the {\sl lateral} shower profile. In this figure, the experimental data are shown in the left ($a$) diagram, and the results of GEANT4 Monte Carlo simulations in the right ($b$) diagram. The fractional differences between the experimental and simulated profiles are shown in Figure \ref{emlprofiles}c.
In Figure \ref{emrprofiles}a, the signals from individual fibers located in the same $r$-bin (\eg 2 - 3 mm from the shower axis) are summed, and the average value of these summed signals is plotted as a function of $r$. We call this the {\sl radial} shower profile. The integral of this profile is normalized to 45\%, \ie the fraction of the shower energy deposited in this calorimeter module. Figure \ref{emrprofiles}b, derived from the same experimental data, illustrates what fraction of the total shower energy is deposited within a certain distance from the shower axis. 

These figures show a remarkable difference between the profiles measured by the two types of fibers. The \v{C}erenkov light is much less concentrated in and near the central fiber than the scintillation light. This is a consequence of the fact that the early, extremely collimated component of the developing shower does not contribute to the \v{C}erenkov signals, since the \v{C}erenkov light falls outside the numerical aperture of the fibers. The consequences of this phenomenon were earlier observed in the muon signals \cite{DREAMmu} and in the angular dependence of the electromagnetic resolution of the RD52 dual-readout fiber calorimeter \cite{RD52_em}.
The figures also show that the GEANT4 Monte Carlo simulations confirmed the substantial difference observed between the profiles measured with the scintillation and \v{C}erenkov signals in great detail. 

The fact that a large fraction of the shower signal comes from only one fiber is already clear from Figure \ref{Scilight}b, which shows that at 10 GeV, this fiber carried 25\% of the total recorded signal. Combined
with the results of the Monte Carlo simulations (Figure \ref{simul}), this means that 10\% of the entire shower energy was deposited within one mm from the fiber axis and contributed to the signal of only one fiber (see also Figure \ref{emrprofiles}b).

\subsection{Caveat}

All experimental results presented in this section are based on the assumption that 45\% of the energy carried by the beam particles was deposited in the tested calorimeter module. This number was provided by GEANT4 Monte Carlo simulations. Implicit in our analyses was the assumption that, therefore, the measured light signals represented 45\% of the light that would be produced in a calorimeter that was large enough to contain the entire shower. 
The light yield in the two types of fibers, the contribution of crosstalk to the \v{C}erenkov signals and the measured shower profiles were all
determined on the basis of this assumption.

However, we have reason to believe that this premise is not entirely correct for the \v{C}erenkov signals. 
In the Appendix, we present the reasons for this believe and derive an alternative fraction based on this. We also calculate the consequences for the measured  
light yields, for the contribution of crosstalk to the \v{C}erenkov signals and for the measured shower profiles.
 
\section{Conclusions}
\vskip -5mm
This is the first time that a dual-readout calorimeter of the type developed in the DREAM and RD52 projects has been equipped with SiPM readout and tested in particle (high-energy electron and muon) beams. The most salient results of these tests are summarized below.
\begin{itemize}
\item The difference in light yield between the scintillation and \v{C}erenkov signals, about a factor of 60 in the number of photoelectrons per GeV deposited energy, is a very challenging feature of this calorimeter. The large scintillation light yield introduces signal saturation effects, already at the sub-GeV level, while the low \v{C}erenkov light yield is responsible for a major contribution to the em energy resolution.
\item An important consequence of the large difference in light yield is the fact that the calorimeter is prone to optical crosstalk effects. Even though the design was primarily inspired by the need to limit/eliminate this crosstalk as much as possible, the \v{C}erenkov signals did contain contributions from scintillation light at the 10 - 20\% level. This will need to be further reduced for a successful application of this calorimeter in experiments that require high-precision energy measurements, since the two types of signals measure different, complementary aspects of the shower development and this information is crucial for a correct measurement of the deposited (hadronic) energy.
\item The fact that every fiber was read out separately made it possible to measure electromagnetic shower profiles with unprecedented precision.
Whereas it is commonly assumed that the radial shower profile scales with the Moli\`ere radius (which is 31 mm in this particular detector), we found that more than half of the shower energy was deposited within 6.5 mm from the shower axis, and 10\% was even deposited within 1 mm. The measurements also confirmed the large differences between the shower profiles measured with scintillation and with \v{C}erenkov light. Hints of this difference were earlier observed in the muon signals and in the angular dependence of the em energy resolution.
\end{itemize}

The main purpose of the project of which this paper is the first report is to find a readout method that would make the dual-readout fiber calorimeter suitable for use in a $4\pi$ collider experiment. This requires that the sensors that convert the light signals into electric pulses be mounted as close to the calorimeter as possible and occupy as little space as possible. SiPMs offer that possibility, provided that they can be organized in such a way that the readout unit fits within the cross-sectional detector area from which the signals are collected. The chosen solution of two arrays of SiPMs looked promising. However, to implement this solution, we depended on what was available on the market. The arrays used in these tests were of course not ideal, in terms of the (square) shape of the SiPMs and the (relatively large) size of the pixels. But it was the {\sl only}  thing available at the time we decided to try this option. In the future, we expect more suitable arrays to be available.

Such arrays should have smaller pixels, which would lead to a better linearity. This is especially important for the SiPMs that detect the scintillation signals. For the \v{C}erenkov signals, an increased quantum efficiency in the blue/UV region of the optical spectrum would increase the PDE, which would directly benefit the achievable energy resolution. Ideally, the SiPMs should also be round, just as the fibers, with a slightly larger diameter. 

In the next stage of this project, we also plan to make a few additional changes, apart from using more optimized SiPM arrays. 
It is clear that a large difference in light yield between the two signals presents a great challenge. In tests with the DREAM and RD52 calorimeters, yellow filters were used for the scintillating fibers. These filters selectively absorb the blue component of the scintillation light. This component is prone to self-absorption and was measured to dominate the light attenuation in these fibers. By equipping the downstream ends of the scintillating fibers with such filters, we expect to reduce the overall light yield by about a factor of five and to increase the light attenuation length substantially \cite{hartjes}. We are also planning to aluminize the upstream ends of the \v{C}erenkov fibers. Based on our previous experience with these techniques, we expect that this will increase the \v{C}erenkov light yield with at least 50\%, thus further reducing the difference between the light yield in the two types of fibers. We believe that these modifications will make it possible  to reduce this difference from the factor of 60 measured in the present tests by an order of magnitude. 

Of course, a next stage should also involve larger calorimeter modules, sufficiently large to contain em showers to the point that leakage fluctuations are negligibly small compared to the envisaged energy resolution. Our simulations have indicated that lateral containment at the 90\% level is needed for 1\% em energy resolution. Such a containment level requires an effective module radius of at least 1.6 $R_M$, or 50 mm, seven times larger than the module of which the test results are described in this paper.

\section*{Acknowledgments}
\vskip-3mm
We thank CERN for making good particle beams available to our experiments in the H8 beam. 
We also thank the team at {\sl Nuclear Instruments} for the fruitful partnership during the development of our data acquisition system. 
This study was carried out with financial support of the United States
Department of Energy, under contract DE-FG02-12ER41783, of Italy's Istituto Nazionale di Fisica Nucleare and Ministero dell'Istruzione, dell'Universit\`a e della Ricerca, and of the Basic Science Research Program of the National Research Foundation of Korea (NRF), funded by the Ministry of  Science, ICT \& Future Planning under contract 2015R1C1A1A02036477.

\section*{Appendix}

The measured shower profiles shown in Figures \ref{emlprofiles}a and \ref{emrprofiles}a indicate that the em showers were less efficiently sampled
by means of the \v{C}erenkov signals, compared to the scintillation signals in our small calorimeter. Nevertheless, we have assumed that the shower containment was the same for both signals (45\%) and Figure \ref{emrprofiles} was made based on that assumption. 
Measurements made with dual-readout calorimeters that were laterally much larger than this one showed no fundamental containment differences. For example, in DREAM \cite{DREAMem}, which had a cell size with a radius of 1.8$R_M$, the lateral containment was measured to be 92\% for the scintillation signals and 93\% for the \v{C}erenkov ones, and in RD52 \cite{RD52_em}, where the cell size had an effective radius of 0.9$R_M$, the lateral containment was 85\%, both for the scintillation and the \v{C}erenkov signals.
These results were based on the measured energy sharing between the central detector cell and the surrounding ones.

However, it may well be that these results were a consequence of the conspiracy of two effects which both tend to reduce the effective sampling fraction measured with the \v{C}erenkov signals:
\begin{enumerate}
\item The radial tails of em showers are dominated by soft electrons from Compton scattering and photoelectric absorption, which are isotropically distributed with respect to the direction of the incoming beam particle, just like the photoelectrons produced by scintillation light. However, \v{C}erenkov light is only produced by electrons with kinetic energies larger than $\sim 200$ keV ($\beta > 0.67$), and therefore the calorimeter is not fully efficient for detecting the soft electron component in the \v{C}erenkov channel.
\item The early, very collimated component of the shower is not (efficiently) detected by the \v{C}erenkov fibers, because of the directional sensitivity. This is clearly demonstrated by the measurements presented in this paper. This component made a substantial contribution to the signals of our small (radius 0.22$R_M$) calorimeter.
\end{enumerate}
\begin{figure}[b!]
\epsfysize=9cm
\centerline{\epsffile{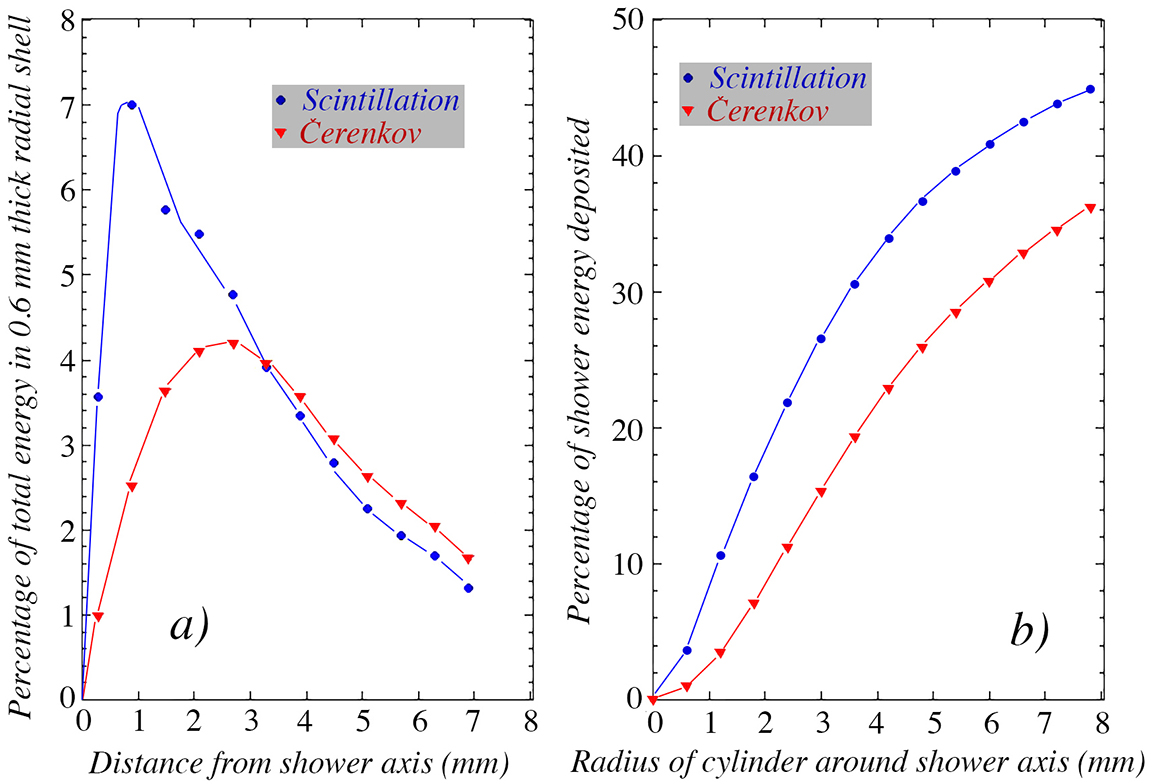}}
\caption{\footnotesize{Radial profiles of electromagnetic showers in the brass-fiber dual-readout calorimeter, measured separately with the \v{C}erenkov
and the scintillation signals ($a$). The fraction of the shower energy deposited in a cylinder around the shower axis as a function of the radius of that cylinder, measured separately with the \v{C}erenkov and the scintillation signals ($b$). These profiles take into account the fact that the containment fraction in this calorimeter is different for the $S$ (45\%) and $C$ (36\%) signals.}}
\label{16mod}
\end{figure}

It may well be that the reduced efficiencies for detecting the \v{C}erenkov component of the light produced in the absorption of em showers near the shower axis and in the radial tails had (approximately) the same net effect on the overall calorimeter signals, thus explaining the containment results mentioned above. However, in our small SiPM calorimeter, only the second effect played a role, and the result could well be a smaller effective sampling fraction than that obtained for the scintillation signals.

To check the possible effects of the mentioned inefficiencies in the shower sampling on our conclusions, we performed additional GEANT4 Monte Carlo simulations, this time of a calorimeter that was large enough to contain the em showers. In these simulations, the signals from the two types of fibers were determined as well. This made it possible to determine what fraction of the total signal was recorded in our small calorimeter.
The results showed that the scintillation signal was in excellent agreement with the expected value based on the energy deposit by the 10 GeV electrons used for these simulations: 45\% of the total signal came from the area covered by our calorimeter. However, for the \v{C}erenkov signals, our calorimeter only accounted for 36\% of the total.

This means that the results presented in Section 4 would have to be revised for calorimeters that are large enough to fully contain em showers.
The \v{C}erenkov light yield was underestimated because it was assumed that the incomplete shower containment in our small calorimeter had the same effect for both types of signals. Using the results mentioned above, the measured signals in the SiPMs connected to the \v{C}erenkov fibers (28.6 photoelectrons per GeV beam energy), would correspond to 28.6/0.36 = 79 photoelectrons per GeV in a sufficiently large calorimeter. After correcting for the crosstalk contribution
(10$\pm$4 photoelectrons), the \v{C}erenkov light yield in such a calorimeter would thus be 69$\pm$5 Cpe/GeV, and the $S/C$ ratio 46$\pm$6.
Figure \ref{16mod} shows the effects of these modifications on the radial shower containment.

\bibliographystyle{unsrt}

\end{document}